\documentclass[aps,twocolumn,showpacs]{revtex4}
\usepackage{graphicx}
\usepackage{bm}
\usepackage{amsmath}

\begin{document}

\title{Quantum Spin Hall Effect in Inverted Type II Semiconductors}
\author{Chaoxing Liu$^{1,2}$, Taylor L. Hughes$^{2}$, Xiao-Liang Qi$^{2}$, Kang Wang$^{3}$ and Shou-Cheng Zhang$^{2}$}

\affiliation{$^1$ Center for Advanced Study, Tsinghua
University,Beijing, 100084, China} \affiliation{$^2$ Department of
Physics, McCullough Building, Stanford University, Stanford, CA
94305-4045} \affiliation{$^3$ Department of Electrical
Engineering, UCLA, Los Angeles, CA 90095-1594}

\begin{abstract}
The quantum spin Hall (QSH) state is a topologically non-trivial
state of quantum matter which preserves time-reversal symmetry; it
has an energy gap in the bulk, but topologically robust gapless
states at the edge. Recently, this novel effect has been predicted
and observed in HgTe quantum wells\cite{bernevig2006d,konig2007}. In
this work we predict a similar effect arising in Type-II
semiconductor quantum wells made from InAs/GaSb/AlSb. Because of a
rare band alignment the quantum well band structure exhibits an
``inverted" phase similar to CdTe/HgTe quantum wells, which is a QSH
state when the Fermi level lies inside the gap. Due to the
asymmetric structure of this quantum well, the effects of inversion
symmetry breaking and inter-layer charge transfer are essential. By
standard self-consistent calculations, we show that the QSH state
persists when these corrections are included, and a quantum phase
transition between the normal insulator and the QSH phase can be
electrically tuned by the gate voltage.

\end{abstract}


\maketitle

Recently, a striking prediction of a quantum spin Hall (QSH)
insulator phase in HgTe/CdTe quantum wells\cite{bernevig2006d} was
confirmed in transport experiments\cite{konig2007}. The QSH
insulator phase is a topologically non-trivial state of matter
reminiscent of the integer quantum Hall effect, but where
time-reversal symmetry is preserved instead of being broken by the
large magnetic field. The state is characterized by a bulk
charge-excitation gap and topologically protected helical edge
states, where states of opposite spin counter-propagate on each
edge\cite{kane2005a,wu2006,xu2006}.Unfortunately, high-quality
HgTe/CdTe quantum wells are very special, and only a few academic
research groups have the precise material control needed to carry
out such delicate experiments. We are therefore lead to search for
other, more conventional, materials that exhibit the QSH effect.

In this work we introduce a new material with the QSH phase, the
InAs/GaSb/AlSb Type-II semiconductor quantum well in the inverted
regime\cite{Chang1980,Altarelli1983,Yang1997,Lakrimi1997,Cooper1998,Halvorsen2000}.
We will show that this quantum well exhibits a subband inversion
transition as a function of layer thickness, similar to the
HgTe/CdTe system, and can be characterized by an effective
four-band model near the transition. This model is similar to the
model for HgTe/CdTe\cite{bernevig2006d}, but contains terms
describing the strong bulk inversion asymmetry (BIA) and
structural inversion asymmetry (SIA). In fact, due to the unique
band alignment of InAs/GaSb/AlSb, the electron subband and the
hole subband are localized in \emph{different} quantum well
layers. Additionally, the band alignment forces one to consider
self-consistent
corrections\cite{Lapushkin2004,Semenikhin2007,Naveh1996} which we
will discuss below. Our results show that the asymmetric quantum
well, with strong built-in electric field, can be electrically
tuned through the phase transition using front and back gates.
While this is of significant fundamental interest, it also allows
one to construct a quantum spin Hall field effect transistor (FET)
that exhibits an insulating ``OFF" state with no leakage current,
and a nearly dissipationless ``ON" state with non-zero conductance
via the topological edge states.

\begin{figure}[htpb]
    \begin{center}
        \includegraphics[scale=0.4]{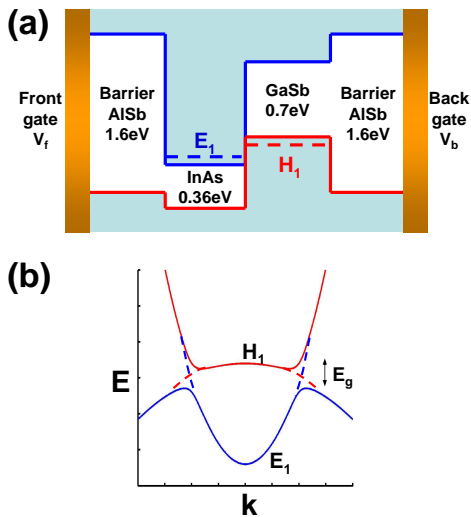}
    \end{center}
    \caption{ (a) Band gap and band offset diagram for asymmetric AlSb/InAs/GaSb quantum wells.
    The left AlSb barrier layer is connected to a front gate while the right
    barrier is connected to a back gate. The $E_1$ subband is localized in the InAs layer and $H_1$ is
    localized in the GaSb layer. Outer AlSb barriers provide an
    overall confining potential for electron and hole states.
     (b) Schematic band structure diagram. The dashed line shows
    the crossing of the $E_1$ and $H_1$ states in the inverted regime, and due to the
    hybridization between $E_1$ and $H_1$, the gap $E_g$ appears.}
    \label{fig:bandedge}
\end{figure}

The quantum well structures in which we are interested are
asymmetric with AlSb/InAs/GaSb/AlSb layers grown as shown in Fig.
\ref{fig:bandedge}. This is an unusual quantum well system due to
the alignment of the conduction and valence band edges of InAs and
GaSb.  The valence band edge of GaSb is $0.15$ eV \emph{higher}
than the conduction band edge of the InAs layer. The AlSb layers
serve as confining outer barriers. The ``conduction" subbands are
localized in the InAs layer while the ``valence" subbands are
localized in the GaSb layer as illustrated in Fig.
\ref{fig:bandedge} (a). In this work we will focus on the regime
where the lowest electron and hole subbands $E_1,H_1$, which are
derived from the $s$-like conduction and $p$-like heavy-hole bands
respectively, are nearly degenerate, and all other subbands are
well-separated in energy. When the quantum well thickness is
increased the energy of the $E_1$ ($H_1$) band edge is decreasing
(increasing). At some critical thickness a level crossing occurs
between $E_1$ and $H_1$, after which the band edge of $E_1$ sinks
below that of $H_1$, putting the system into the inverted regime
of Type-II quantum wells. Since the $H_1$ band disperses downwards
and the $E_1$ band disperses upwards, the inversion of the band
sequence leads to a crossing of the two bands, see Fig.
\ref{fig:bandedge} (b). Historically, the inverted regime of
InAs/GaSb/AlSb quantum wells was described as a semi-metal without
a gap\cite{Chang1980}. However, Ref. \cite{Altarelli1983} first
pointed out that due to the mixing between $E_1$ and $H_1$, a
small gap ($E_g$ in Fig \ref{fig:bandedge} (b)) is generally
opened, leading to bulk insulating behavior. This hybridization
gap was later demonstrated in
experiments\cite{Yang1997,Lakrimi1997}. Therefore, just like in
the HgTe/CdTe quantum wells, the inverted regime of InAs/GaSb
quantum wells should be a topologically non-trivial QSH phase
protected by the bulk gap.

This seemingly simple conclusion is complicated by the unique
features of type II quantum wells: the electron-subband and
hole-subband are separated in two different layers. There are
several separate consequences of this fact. First, the hybridization
between $E_1$ and $H_1$ is reduced, but this is just a quantitative
correction. Second, since there is no inversion symmetry in the
quantum well growth direction, SIA terms may be large enough to
compete with the reduced hybridization. In addition, BIA may also
play a role for this system. Therefore, both SIA and BIA must be
included properly to make a correct prediction, while in HgTe/CdTe
quantum wells these two types of terms were ignored because BIA
terms are small when compared with the hybridization, and the
quantum well was symmetric which minimizes SIA. Finally, since the
electron and hole subbands lie in two different layers, there is an
automatic charge transfer between the layers which yields a
coexistence of p-type and n-type carriers. Consequently, a
self-consistent treatment of Coloumb energy is necessary to account
for this effect. In the following, we will discuss all of these
issues and conclude that the QSH phase exists in an experimentally
viable parameter range.

The materials in these quantum wells have the zinc-blende lattice
structure and direct gaps near the $\Gamma$ point and are thus
well-described by the $8$-band Kane model\cite{kane1957}. We will
construct an effective $4$-band model using the same envelope
function approximation procedure as the Bernevig-Hughes-Zhang
(BHZ) model\cite{bernevig2006d}; albeit a more complex one due to
the SIA and BIA terms. The Hamiltonian naturally separates into
three distinct parts
\begin{equation}
{\cal H}=H_0+H_{BIA}+H_{SIA}\label{Ham}.
\end{equation}\noindent In the basis $\{ \vert E_1 +\rangle,\vert H_1 +\rangle,\vert E_1 -\rangle,\vert H_1
-\rangle \},$ and keeping terms only up to quadratic powers of
$\textbf{k},$ we have
\begin{eqnarray}
H_0=\epsilon(k)\mathbf{I}_{4\times
4}+\left(\begin{array}{cccc}\mathcal{M}(k)&Ak_+&0&0\\Ak_-&
-\mathcal{M}(k)&0&0\\0&0&\mathcal{M}(k)&-Ak_-\\0&0&-Ak_+&
-\mathcal{M}(k)\end{array}\right) \label{Ham_0}
\end{eqnarray}\noindent where $\mathbf{I}_{4\times 4}$ is the $4\times 4$ identity
matrix, $\mathcal{M}(k)=M_0+M_2k^2$ and $\epsilon(k)=C_0+C_2k^2$.
This is simply the Hamiltonian used by BHZ. The zinc-blende
structure has two different atoms in each unit cell, which breaks
the bulk inversion symmetry and leads to additional terms in the
bulk Hamiltonian\cite{winkler2003}. When projected onto the lowest
subbands the BIA terms are
\begin{eqnarray}
    H_{BIA}=\left(\begin{array}{cccc}0&0&\Delta_ek_+&-\Delta_0\\0&0&\Delta_0&\Delta_hk_-\\
    \Delta_ek_-&\Delta_0&0&0\\-\Delta_0&\Delta_hk_+&0&0\end{array}\right).
    \label{Heff_BIA}
\end{eqnarray}
Finally the SIA term reads
\begin{eqnarray}
    H_{SIA}=\left(\begin{array}{cccc}0&0&i\xi_ek_-&0\\0&0&0&0\\
    -i\xi_e^*k_+&0&0&0\\0&0&0&0\end{array}\right).
    \label{Heff_SIA}
\end{eqnarray}\noindent
Here we recognize the SIA term as the electron $k$-linear Rashba
term; the heavy-hole $k$-cubic Rashba term is neglected. The
parameters $\Delta_h,\Delta_e,\Delta_0,\xi_e$ depend on the
quantum well geometry.

Now we address, from pure band structure considerations, whether
or not a QSH phase exists in this model. Without $H_{BIA}$ and
$H_{SIA}$ the Hamiltonian is block diagonal and each block is
exactly a massive Dirac Hamiltonian in $(2+1)d$. By itself, each
block breaks time-reversal symmetry, but the two $2\times2$ blocks
are time-reversal partners so that the combined system remains
time-reversal invariant. As mentioned, this is the pure BHZ model
and from their argument we know that there is a topological phase
transition signalled by the gap closing condition $M_0=0$, and the
system is in QSH phase when $M_0/M_2<0$. When $H_{BIA}$ and
$H_{SIA}$ terms are included, the two blocks of $H_0$ are coupled
together and the analysis in BHZ model does not directly apply.
However, the QSH phase is a topological phase of matter protected
by the band gap\cite{kane2005a,wu2006,xu2006}. In other words, if
we start from the Hamiltonian $H_0$ in the QSH phase and turn on
$H_{BIA}$ and $H_{SIA}$ adiabatically, the system will remain in
the QSH phase as long as the energy gap between $E_1$ and $H_1$
remains finite. With realistic parameters for an InAs/GaSb/AlSb
quantum well obtained from the $8$-band Kane model, the adiabatic
connection between the inversion-symmetric Hamiltonian $H_0$ and
the full Hamiltonian $\mathcal{H}$ was verified for the proper
parameter regime, which supports the existence of a QSH phase in
this system. Though the BIA and SIA terms do not destroy the QSH
phase, they do modify the quantum phase transition between the QSH
phase and normal insulator (NI). The transition (gap-closing) will
generically occur at finite-${\bf k}$ rather than at the $\Gamma$
point, and a nodal region between QSH and NI phases can possibly
appear in the phase
diagram\cite{murakami2007,Dai2008,koenig2008,hughes2007}.

\begin{figure}[t!]
    \begin{center}
        \includegraphics[width=2.7in]{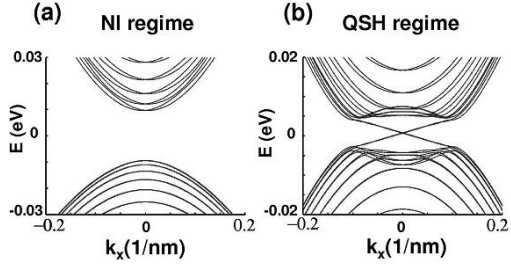}
    \end{center}
    \caption{The energy dispersions of Hamiltonian (\ref{Ham}) on a cylindrical geometry with open boundary conditions
    along the $y$-direction and periodic boundary conditions along
    the $x$-direction. (a) The dispersion for the quantum well with
    GaSb layer thickness $d_1=10{\rm nm}$ and InAs layer thickness $d_2=8.1{\rm nm}$, which is a normal insulator
    with no edge states. (b) The dispersion for
    $d_1=d_2=10{\rm nm}$ quantum well which is a QSH insulator with one pair of edge states.
    A tight-binding regularization with lattice constant $a=20\AA$ is used in this calculation.}
    \label{fig:edge}
\end{figure}

A more direct way of identifying the QSH phase is to study the
edge state spectrum. There are always an odd number of Kramers's
pairs of edge states confined on the boundary of a QSH insulator,
and an even number pairs (possibly zero) for the boundary of the
NI phase. The edge state energy spectrum of the effective model
(\ref{Ham}) can be obtained by solving this model with a simple
tight-binding regularization in a cylindrical geometry, the result
of which is shown in Fig. \ref{fig:edge}. We find one Kramers's
pair of edge states with opposite spin on each edge for the QSH
side, and no edge states for the NI side. This again confirms the
existence of QSH phase in this model.

\begin{figure}[t!]
    \begin{center}
        \includegraphics[width=2.4in]{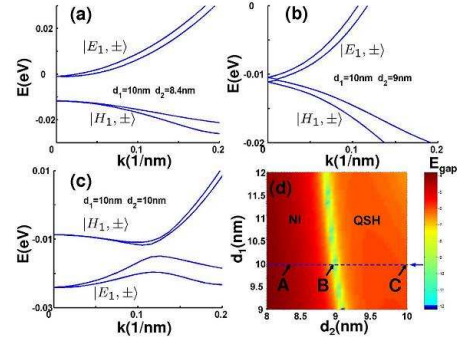}
    \end{center}
    \caption{(a)-(c) The energy dispersions calculated from the $8$-band
    Kane model for three well configurations, where
    $d_1$ and $d_2$ are the thickness of GaSb layer and InAs layer, respectively.
    (d) The energy gap variation in $d_1-d_2$ plane, where brighter colors represent a smaller gap.
    A, B and C on the dashed blue line indicate respectively
    the place where (a), (b) and (c) are plotted. NI and QSH denote the
    phases in the corresponding region of parameter space. }
    \label{fig:Dispersion}
\end{figure}

To study the InAs/GaSb/AlSb quantum well system more
systematically and quantitatively, we confirm the above analysis
by numerically solving the realistic $8$-band Kane model. In the
inverted regime, there exists an intrinsic charge transfer between
the InAs layer and GaSb layer. Therefore, we need to take into
account the built-in electric field. The energy dispersions for
different well thicknesses are shown in Fig. \ref{fig:Dispersion}
(a)-(c), where we fix the GaSb layer thickness $d_1=10{\rm nm}$
and vary the thickness of InAs layer $d_2$. The system is gapped
for a generic value of $d_2$. However, at a critical thickness
$d_{2c}=9{\rm nm}$ (Fig \ref{fig:Dispersion} (b)) a crossing at
finite ${\bf k}$ occurs between the subbands $E_1$ and $H_1$,
which marks the phase transition point between the QSH and NI
phases. According to the above adiabatic continuity argument, we
know that the quantum well is in a NI state for $d_2<d_{2c}$ (Fig.
\ref{fig:Dispersion} (a)) and QSH state for $d_2>d_{2c}$ (Fig.
\ref{fig:Dispersion} (c)). As the band inversion is only
determined by the relative positions of $E_1$ and $H_1$, the
quantum wells with other values of $d_1$ behave essentially the
same. As the QSH phase and NI phase are always separated by a gap
closing point, we can determine the $d_1-d_2$ phase diagram via
the energy gap. As shown in Fig. \ref{fig:Dispersion} (d), two
gapped regimes (in red) are separated by a critical line (brightly
colored) in the $d_1,d_2$ plane. The quantum well configurations
shown in Figs. \ref{fig:Dispersion} (a), (b) and (c) are indicated
by points A, B and C, respectively. Due to the adiabatic
continuity, an entire connected gapped region in the phase diagram
is in the NI (QSH) phase once one point in it is confirmed to be
in this phase. Since Fig. \ref{fig:Dispersion} (a) corresponds to
the NI phase and (c) the QSH phase, we identify the right side of
the diagram as the QSH regime and the left side as the NI regime.

One advantage of the InAs/GaSb/AlSb quantum wells is that due to
the large built-in electric field, the QSH-NI phase transition can
be easily tuned by external gate voltages. When we tune the gate
voltage, both the band structure and the Fermi level are adjusted
simultaneously. Since the QSH effect can only occur when the Fermi
level lies in the gap, we need two gates in order to independently
tune the relative position between the $E_1$ and $H_1$ band edges
and the Fermi level. In fact, such a dual-gate geometry has
already been realized experimentally in InAs/GaSb/AlSb quantum
wells\cite{Cooper1998}. In the present work, we performed a
self-consistent Poisson-Schrodinger type
calculation\cite{Lapushkin2004,Semenikhin2007,Naveh1996} for such
a dual-gate geometry shown in Fig. \ref{fig:bandedge} (a).
To simplify the calculation, we take the thickness of the AlSb
barrier layers to be much smaller than that in realistic
experiments, which has a negligible effect in the quantum well
except for a rescaling of $V_f$ and $V_b$. We also neglect the
weak effects of subband anisotropy and intrinsic donor defects at
the InAs/GaSb interface. None of these simplifications should
affect our results qualitatively.

\begin{figure}[htpb]
    \begin{center}
        \includegraphics[width=3in]{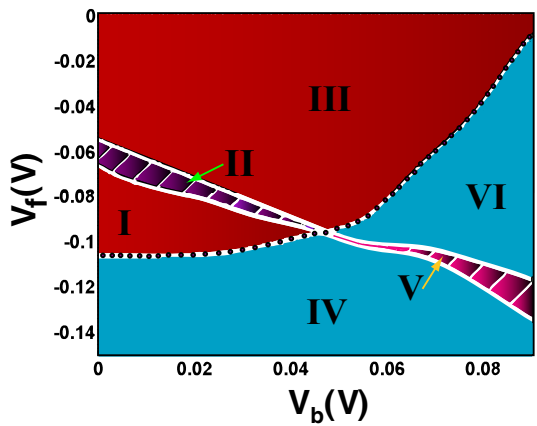}
    \end{center}
    \caption{The phase diagram for different front ($V_f$) and back ($V_b$) gate voltages.
   Regions I,II,III are in the inverted regime, in which the striped region II is the
   QSH phase with Fermi-level in the bulk gap, and I, III are the p-doped and n-doped inverted system.
   Regions IV,V,VI are in the normal regime, in which the striped region V is the
   NI phase with Fermi level in the bulk gap, and IV, VI are the
   p-doped and n-doped normal semiconductors. The well configuration is set as $d_1=d_2=10nm$, and the
   AlSb barrier thickness is taken 30nm on each side in the self-consistent
   calculation. $V_f$ and $V_b$ are defined with respect to the Fermi level in the quantum well. }
    \label{fig:selfconsistent}
\end{figure}

For fixed $d_1=d_2=10$ nm we explored the $V_f-V_b$ phase diagram
as shown in Fig. \ref{fig:selfconsistent}. There are six distinct
regions in the figure. The dotted black line shows the gap closing
transition between the inverted and non-inverted regimes. In
parameter regions I,II,III the system has an inverted band
structure, but only region II is in the QSH phase with the Fermi
level tuned inside the bulk gap. Region I (III) is described by
the same Hamiltonian as the QSH phase, but with finite hole
(electron) doping. In the same way, region V is the NI phase and
IV, VI are the corresponding p-doped and n-doped normal
semiconductors. Thus, by tuning $V_f$ and $V_b$ to the correct
range, one can easily get the phase transition between QSH phase
II and NI phase V.

Compared to the similar proposal of a gate-induced phase
transition in asymmetric HgTe/CdTe quantum wells\cite{Yang2007},
the InAs/GaSb/AlSb quantum well is much more sensitive to the gate
voltage, which makes it much easier to realize such a transition
experimentally. Physically, this comes from the fact that the
electron and hole wavefunctions are centered in separate layers,
so that the effect of the gate voltage on them is highly
asymmetric. This simple mechanism allows us to investigate the
quantum phase transition from the NI to the QSH state in-situ,
through the continuous variation of the gate voltage, rather than
the discrete variation of the quantum well thickness. It is also
useful for developing a QSH FET. The FET is in an `OFF' state when
the Fermi level lies inside the normal insulating gap. Then, by
adjusting the gate voltages the FET can be flipped to the `ON'
state by passing through the transition to the QSH phase, where
the current is carried only by the dissipationless edge states.
This simple device can be operated with reasonable voltages as
seen in Fig. \ref{fig:selfconsistent} but would be more promising
if one could enlarge the bulk insulating gap to support room
temperature operation.

In conclusion, we propose that the QSH state can be realized in
InAs/GaSb quantum wells. We presented both simple arguments based
on effective model and realistic self-consistent calculations. In
addition we have proposed an experimental setup to electrically
control the quantum phase transition from the normal insulator to
the QSH phase. This principle could be used to construct a QSH FET
device with minimal dissipation.

We would like to thank B. F. Zhu for the useful discussion. This
work is supported by the NSF under grant numbers DMR-0342832, the US
Department of Energy, Office of Basic Energy Sciences under contract
DE-AC03-76SF00515 and by the Focus Center Research Program (FCRP)
Center on Functional Engineered Nanoarchitectonics (FENA). CXL
acknowledges the support of China Scholarship Council, the NSF of
China (Grant No.10774086, 10574076), and the Program of Basic
Research Development of China (Grant No. 2006CB921500).

\end{document}